\documentclass[aps,prl,reprint,groupedaddress]{revtex4-1}

\usepackage{amssymb,amsfonts,amsmath}
\usepackage{graphicx}
\usepackage{dcolumn}
\usepackage{bm}
\usepackage{color}
\usepackage{epstopdf}

\begin{document}
\newcommand{\comment}[1]{}
\newcommand{\Orf}{\Omega_\textrm{rf}}
\newcommand{\Sr}{$^{88}$Sr$^+$ }
\newcommand{\kB}{\textrm{k}_\textrm{B}}
\newcommand{\ket}[1]{\left| #1 \right>} 
\newcommand{\bra}[1]{\left< #1 \right|} 
\newcommand{\braket}[1]{\left< #1 \right>} 
\newcommand{\p}[2]{\left|#1\left>\right<#2\right|} 
\newcommand{\level}[3]{\textrm{#1}_{#2}\left(#3\right)}

\title{Single-shot energy measurement of a single atom and the direct reconstruction of its energy distribution}

\author{Ziv Meir}
\author{Tomas Sikorsky}
\author{Nitzan Akerman}
\author{Ruti Ben-shlomi}
\author{Meirav Pinkas}
\author{Roee Ozeri}
\affiliation{Department of Physics of Complex Systems, Weizmann Institute of Science, Rehovot 7610001, Israel}
\date{\today}

\begin{abstract}
An ensemble of atoms in steady-state, whether in thermal equilibrium or not, has a well defined energy distribution. 
Since the energy of single atoms within the ensemble cannot be individually measured, energy distributions are typically inferred from statistical averages. 
Here, we show how to measure the energy of a single atom in a single experimental realization (single-shot).
The energy distribution of the atom over many experimental realizations can thus be readily and directly obtained.     	
We apply this method to a single-ion trapped in a linear Paul trap for which energy measurement in a single-shot is applicable from 10 K$\cdot\kB$ and above.
Our energy measurement agrees within 5$\%$ to a different thermometry method which requires extensive averaging.
Apart from the total energy, we also show that the motion of the ion in different trap modes can be distinguished.  
We believe that this method will have profound implications on single particle chemistry and collision experiments.  
\end{abstract}

\maketitle



\section{\label{sec:introduction}Introduction}
Thermometry is a fundamental tool in the natural sciences \cite{Moldover2016}. 
A single-atom system can be classically assigned with a well-defined energy at any given moment in time. 
Temperature or, more generally, energy distribution arises when the energy is recorded over many identical experimental realizations \cite{Leibfried2003}.
Typically, a reliable measurement of a single atom's energy in a single experimental realization is not possible. Hence, the energy distribution is reconstructed from the average over multiple experimental realizations. The resulting signal is then analyzed using the assumed underlying distribution parameters as free fit parameters \cite{Wesenberg2007, Meir2016, Sikorsky2017}.

Here, we propose and implement a method to directly measure the energy of a single trapped atom in a single experimental realization. We detect the atom's fluorescence during the process of laser Doppler-cooling from which we extract the atom's energy in a "single-shot". The energy distribution of the atom over multiple realizations is thus directly measured without any prior assumptions.
Doppler-cooling thermometry \cite{Wesenberg2007,Sikorsky2017} is a well-known technique for trapped atoms in the energy regime starting from the Doppler temperature limit in the mK range up to a few K. Here, we extend this method to a regime of higher energies in which the high non-linearity of the fluorescence signal in time allows exact determination of the atom's energy in a single measurement. 
The extension of Doppler thermometry to 100's and even 1000's K opens new experimental avenues for the determination and analysis of exothermic processes \cite{WillitschFirst,Hall2011,Harter2012,Ratschbacher2012,Tong2012,Artjum_threebody,Artjum_threebody2,Sikorsky2017sr,Benshlomi2017} and out-of-equilibrium dynamics \cite{BlueSky,Meir2017neq}. 

Other methods for reconstructing the energy (and/or velocity) distribution like velocity-mapping-imaging (VMI) \cite{eppink1997velocity} rely on photo-ionization or detachment of the atom from one of its electrons for detection. This makes these type of methods destructive to the initial state of the atom as they are usually used to analyze the ionization processes itself. The method presented here does not involve destructive state-changing processes, and the atom is not lost after detection.

We validate our energy measurement accuracy by comparing it to a different thermometry technique; namely fluorescence imaging of the ion oscillation amplitude in the trap; which relies on extensive averaging and performed in a lower energy regime. We found that both techniques agree to within 5$\%$. As an application of our method, we use it to extract the small non-linearity of the Paul trap potential \cite{Akerman2010}. Since the anharmonicity of the trap potential is small, its measurement requires the excitation of the ion to very high energies.

In a three-dimensional trap, the atom's energy is distributed between the different modes of the trapping potential.
In general, and specifically in linear Paul traps, the different modes have a slightly different cooling dynamics which can potentially introduce errors to our energy estimation. 
We show that a likelihood analysis can distinguish between the modes even in a single-shot measurement, which improves the method accuracy in determining the total energy and also its applicability to wider experimental situations. 
This feature of our thermometry can also be used to mark the occurrence of a collision which changes the initial direction of ion motion without changing its energy \cite{Meir2017neq}. 

\section{\label{sec:method}Method}
Our model calculates the trapped particle motion and density matrix of internal states during the process of Doppler cooling by solving the classical equations of motion (EOM - Eq. \ref{eq:EOMx}-\ref{eq:EOMv}) and the time dependent optical Bloch equations (OBE - Eq. \ref{eq:OBE}) simultaneously. These sets of equations are coupled through the Doppler shifts of internal transitions due to motion and the Rabi frequency modulation due to finite size of the laser beams.
This method is applicable for any particle with closed cycling cooling transitions. \cite{Wineland1979,Shuman2010,Lindenfelser2016}.

\begin{figure} 
	\centering
	\includegraphics[width=\linewidth, trim=1.5cm 1cm 12.7cm 10cm, clip]{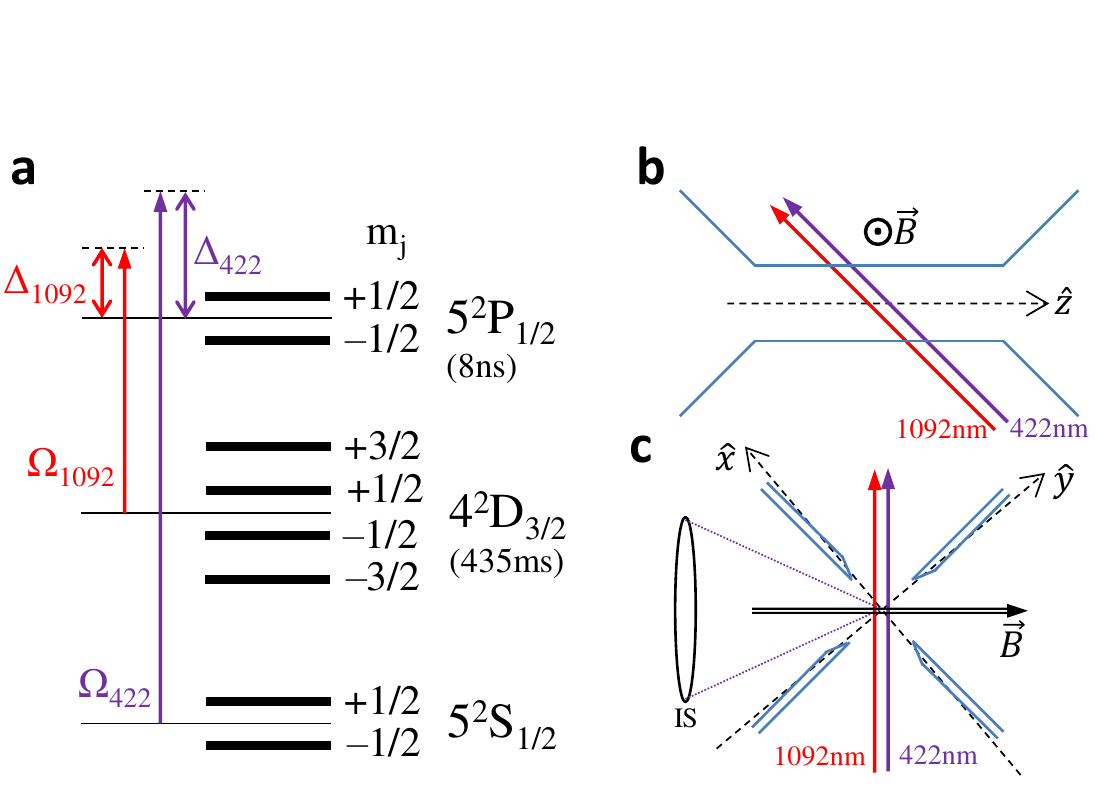}
	\caption{\textbf{a) \Sr energy levels.} The three energy manifolds which participate in Doppler cooling are the $5^2$S$_{1/2}$ ground-state, the $4^2$D$_{3/2}$ meta-stable state and the $5^2$P$_{1/2}$ excited-state. Each manifold splits under magnetic field according to the projection of total angular momentum on the magnetic field axis, $m_j$. Lifetimes of energy levels are given in brackets. Cooling beam (purple) and repump beam (red) are indicated together with their corresponding detuning ($\Delta_{422}/2\pi=-19$  MHz, $\Delta_{1092}/2\pi=0$ MHz) and Rabi-frequencies ($\Omega_{422}/\Gamma_{422}=0.5$ and $\Omega_{1092}/\Gamma_{1092}=8$). Diagram is not to scale. \textbf{b) Experimental setup side-view.} Cooling (purple) and repump (red) beams are co-propagating, 46$^\circ$ with respect to the axial ($\hat{z}$) axis of the ion trap (trap electrodes in blue). \textbf{c) Experimental setup front view.} Beams projection on the ion trap's radial plane is 40$^\circ$ with respect to the radial $\hat{x}$ mode. The magnetic field, $|\vec{B}|=3\pm0.02$ Gauss, quantization axis is perpendicular to the cooling and repump beams. The beams' linear polarization angle (not shown) is 6$^\circ$ and 35$^\circ$ with respect to the magnetic field for the cooling and repump respectively. The imaging system (IS) collects only 422 nm photons. The collection efficiency (photons collected/scattered) is 1/(190$\pm$2). }\label{fig:exp}
\end{figure}

Here, we consider a single \Sr ion trapped in a linear rf Paul trap (for more details on the experimental apparatus see \cite{Meir2017exp}) subject to close-to-resonance cooling (422 nm) and repump (1092 nm) laser beams. The scheme of the eight energy levels that participate in laser cooling is shown in Fig. \ref{fig:exp}a. A sketch of the experimental apparatus is given in Fig. \ref{fig:exp}b-c. Experimental details are given in the figure caption. 

The EOM of the ion are given by,
\begin{align}
	\dot{x}_i=&v_i,\label{eq:EOMx}\\
	\dot{v}_i=&-x_i(a_i+2q_i\cos(\Orf t))\Orf^2/4+\label{eq:EOMv}\\
	&\hbar\rho_e\left(\Gamma_{422} k_{i,422}+\Gamma_{1092} k_{i,1092}\right)/m.\nonumber
\end{align}
Here, $a_i$ and $q_i$ are the Mathieu trap-parameters which determine the magnitude of the Paul trap static and rf quadrupole electric fields respectively \cite{Paul1990}. The rf frequency of the oscillating electric fields is $\Orf$. The second term in Eq. \ref{eq:EOMv} is the effective cooling force generated by the absorption and spontaneous emission of the cooling and repump lasers photons \cite{Wineland1979}. The transitions linewidths are $\Gamma_{422}/2\pi=20.37$ MHz and $\Gamma_{1092}/2\pi=1.18$ MHz and $k_i$ are the laser k-vector projections on the different trap modes ($i=x,y,z$). Planck constant is $\hbar$ and $m$ is the ion's mass. The excited state population is given by $\rho_e=\rho_{33}+\rho_{44}$ where $\rho_{ij}=\left<\psi_i|\hat{\rho}|\psi_j\right>$ and $\left|\psi_{3|4}\right>=\left|\textrm{P}_{1/2},m=\mp1/2\right>$.

To solve the particle's EOM, we need to determine the excited-state population. We write the time evolution of the particle's density matrix, $\hat{\rho}$, in the form of a Lindblad master equation,
\begin{equation}\label{eq:OBE}
\begin{split}
	&\frac{\textrm{d}\hat{\rho}}{\textrm{d}t}=\mathcal{L}{\left[\hat\rho\right]},\\
	&\mathcal{L}=-\frac{i}{\hbar}\left[\hat{H},\hat{\rho}\right]+\hat{D}.
\end{split}
\end{equation}
Here, $\hat{H}$ is the Hamiltonian of the system which is composed of the atomic energy levels (Fig. \ref{fig:exp}a) and the light-matter interaction. $\hat{D}$ is the Lindblad operator which includes dissipative processes such as spontaneous emission (We follow \cite{Oberst1999} in our derivation. Full details can be found in \cite{Sikorsky2017}). 

We solve the set of 70 coupled first-order equations given in Eq. \ref{eq:EOMx}-\ref{eq:OBE} using the Runge-Kutta 4$^\textrm{th}$ order method. We exploit the hermicity of the density matrix to reduce the number of equations in \ref{eq:OBE} from 64 to 32. We check during the numerical integration that the density matrix trace, $\textrm{Tr}\left(\hat\rho\right)=1$, is preserved. 

Low-energy Doppler cooling methods \cite{Wesenberg2007,Sikorsky2017} use effective models to generate the ion's fluorescence signal, which are based on the assumption that the ion internal states reach steady state faster than the ion moves. These methods are numerically much more efficient than the method presented here. However, they are limited to the above assumption and they become impractical at energies above 10's K due to a large number of side-bands associated with the inherent micromotion. 

\begin{figure} 
	\centering
	\includegraphics[width=\linewidth, trim=3.5cm 8.5cm 3.5cm 8.5cm, clip]{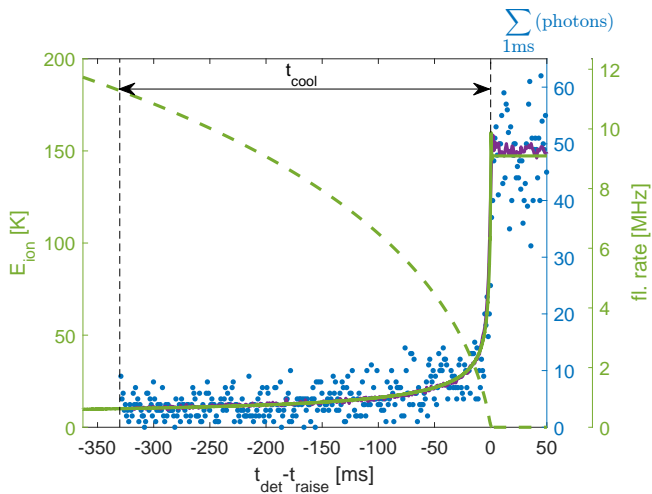}
	\caption{\textbf{Axial mode single-shot energy measurement.} Ion's secular energy (dashed green line) and fluorescence rate (solid green line) derived from the numerical solution of Eq. \ref{eq:EOMx}-\ref{eq:OBE} where the initial energy is only in the axial mode. We add to the numerical results a constant background which we measure independently (1.89$\pm$0.02 photons in 1 ms). Blue points are an example of a single experimental realization measured fluorescence rate using 1 ms time-binning. We find the cooling time using maximum likelihood analysis of the data to the numerical results. From the cooling time, $\textrm t_\textrm{cool}=330.2\pm0.6$ ms, we extract the ion's initial energy, $\textrm E_\textrm{ion}=182.74\pm0.14$ K. Errors accounts only for statistical noise (fit confidence). In this figure, we set the zero time axis to overlap the onset of reaching the Doppler temperature for both the numerical results and experimental data. Purple line is the average over 100 identical experimental realizations.}\label{fig:singlemode}
\end{figure}

Our numerical results are shown in Fig. \ref{fig:singlemode} (green lines). The initial conditions of the ion's amplitude correspond to 225 K in the axial mode while the amplitudes in the other modes are set to zero. We plot the ion's secular energy (dashed green line in Fig. \ref{fig:singlemode}) defined as $\textrm{E}_\textrm{ion}(t)=\sum_i m \omega_i^2 x_{0,i}(t)^2/2$ where $\omega_i$ are the trap secular frequencies and $x_{0,i}$ are the secular amplitudes calculated using the 1$^\textrm{st}$ order solution of the Mathieu equations (Eq. \ref{eq:EOMv} without the damping term) \cite{Paul1990}. We also plot the fluorescence rate defined as $\Gamma_\textrm{ion}(t)=\Gamma_{422}\rho_e(t)$ (solid green line). 

The fluorescence signal features high non-linearity in time. As the ion cools down fluorescence is maintained at a low rate of $\sim350$ kHz 
and almost does not increase for $\sim350$ ms until the ion reaches energy of $\sim25$ K. At this point, the fluorescence rate increases rapidly, within few ms, to the steady-state value of $\sim8.7$ MHz. 
Previous Doppler cooling thermometry \cite{Wesenberg2007,Sikorsky2017} was performed in the latter regime, where the fluorescence signal changes significantly and rapidly during the cooling process. Due to the short time-scale, typically few ms, of the dynamics in this regime, the signal-to-noise ratio is poor and averaging over many experimental realizations is required. In the single-shot Doppler thermometry presented here, we focus on the former regime. We exploit the high non-linearity characterized by the sharp rise in the fluorescence rate at the end of the cooling dynamics to precisely determine the time it takes the ion to cool down. Using our simulation we can translate initial energy to cooling time and vice-versa (see Fig. \ref{fig:singlemode}). We can do so since the cooling process is deterministic and heating of the other modes is negligible.

\section{Single mode experiment}
We compare the numerical results of the fluorescence rate to the one observed in experiment. We trap a single \Sr ion and cool it to the Doppler temperature (0.5 mK). We then increase its kinetic energy significantly in the axial mode using an oscillating electric field pulse close-to-resonance ($\delta=2\pi\cdot150$ Hz) with the axial mode frequency ($\omega_z=418.561(3)$ kHz). We detect the ion's fluorescence using a photon-counter in 1 ms time-bins. For our experimental parameters, this amounts for $\sim3$ photons/bin at the beginning and $\sim50$ photons/bin at the end of the fluorescence signal. A single-shot experimental result is shown in Fig. \ref{fig:singlemode} (blue dots). We use a maximum likelihood analysis to match the experimental cooling time data to the numerical results using a single fit parameter - the ion's initial energy. We estimate an initial energy of $\textrm E_\textrm{ion}\approx183$ K. Our numerical results agree very well with the single-shot experimental data (blue dots) and also with an average over 100 experimental realizations (solid purple line). Details regarding the derivation of the numerical calculation parameters are given in the supplemental material (SM). The statistical noise in determining the cooling time introduces only a small error (see Fig. \ref{fig:singlemode} caption) compared to the error due to the uncertainty in the numerical calculation parameters. We estimate the later error to be on the level of $\sim2\%$ for this energy regime (see SM).  

To test the accuracy of our energy measurement we need to compare our method to a different energy measurement scheme. However, we could not find a method to measure ion energies in the energy range above 10's K. Instead, we compare our high energy result to a low energy thermometry method in the few K regime. We link between the two energy regimes by comparing the fitted experimental parameters of both methods.

In both experiments, we heat the ion using close-to-resonance oscillating electric-field pulse on the trap electrodes. We model heating using a non-damped driven harmonic oscillator EOM together with a cubic non-linear term,
\begin{equation}\label{eq:nonlinear}
	\ddot z + \omega_z^2 z + (\omega_z/z_{nl})^2 z^3 = e E_{d} \cos((\omega_z+\delta)t)/m.
\end{equation}
Here, $z_{nl}$ accounts for the trap non-linearity, $E_{d}$ is the electric field drive amplitude, $\delta$ is the drive frequency detuning from the axial resonance, $\omega_z$. We can treat only a single axis of the ion trap since the detuning of the drive from the other modes is large (100's kHz). We use the harmonic oscillator equation instead of the Mathieu equation (Eq. \ref{eq:EOMv}) since the amplitude of rf fields along the axial direction in our trap is negligible and trapping in this direction is solely due to static fields \cite{Meir2017exp}. 

In the first experiment, we scan the drive pulse time and measure the ion energy using our single-shot energy measurement. We observe periodic oscillation of the ion energy as expected from a driven, non-damped harmonic oscillator \cite{Ozeri2011}. However, the period of this oscillation ($\sim4$ ms) is faster than what we expect from the detuning of our driving field ($1/150$ Hz=6.67 ms). The fast oscillations occur due to the trap non-linearity which becomes important in this energy regime. As the ion oscillation amplitude increases, non-linearity pulls the trap frequency away such that the effective detuning increases.

\begin{figure}
	\centering
	\includegraphics[width=\linewidth, trim=3.5cm 13cm 5cm 8cm, clip]{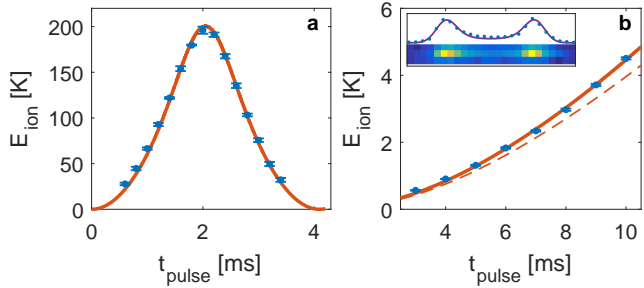}
	\caption{\textbf{Comparison of two thermometry methods.} a) Dots are energies extracted using the single-shot Doppler cooling thermometry. Each data point is the average of six experiments. Error-bars are the standard-deviation of repeated experiments. Line is a fit to  the solution of Eq. \ref{eq:nonlinear} with two fit parameters: $z_{nl}$ and $E_{d,1}$. b) Dots are energies extracted from measuring the ion's amplitude on a CCD (inset). Solid line is a fit with two parameters: $E_{d,2}$ and $\delta$. $E_{d,1}$ and $E_{d,2}$ are related through the ratio of the electric field power between the two experiments. The measured ratio (33.6$\pm$0.6) and the actual ratio (35.5) agrees at the 5$\%$ level. Dashed line is a numerical calculation using the electric fields, $E_{d,2}$, calculated from $E_{d,1}$ which graphically shows the small systematic error between the methods.
	}\label{fig:nonlinear}
\end{figure}

We numerically solve Eq. \ref{eq:nonlinear} with the initial conditions of an ion at rest in the center of the trap. We calculate the total ion energy, $\textrm{E}_\textrm{ion}(t)=m(\omega_z^2 z^2(t) + \dot z^2(t))/2$, as a function of the pulse time. We fit the solution to the experimental results with two fit parameters, the trap non-linearity ($z_{nl}=1146\pm5$ $\mu$m) and the electric field amplitude ($E_{d,1}=197.7\pm1.3$ mV/m). The fit and experimental results are shown in Fig. \ref{fig:nonlinear}a. The electric field value is necessary for the comparison of our thermometry with an independent method as shown below. The non-linearity coefficient is a measure of the Paul trap anharmonicity. To measure this trap parameter, high-energy motional excitation of the ion is required, as in the method presented here. Characterization of the anharmonicity in Paul traps is important, e.g, in the field of mass spectrometry \cite{Makarov1996}.

In the second, low energy, experiment, we also scan the drive pulse time however with a much lower drive amplitude. We measure the ion oscillation amplitude by imaging the spatial distribution of fluorescence on a CCD, from which we can derive the ion's energy (Fig. \ref{fig:nonlinear}b inset). Data was taken from \cite{Sikorsky2017}. Due to the mechanical effects of fluorescence on the ion, we scatter only a few photons in each experimental repetition and repeat 5,000 times to improve our signal-to-noise. Due to the oscillatory motion of the ion, the fluorescence stretches along the axis of motion. We extract the oscillation amplitude from a fit to a model which takes into account the effect of Doppler shifts on the fluorescence (purple line in Fig. inset, see SM for futher details). We fit the solution of Eq. \ref{eq:nonlinear} to the experimental results with two fit parameters, the electric field amplitude ($E_{d,2}=5.89\pm0.09$ mV/m) and the drive detuning ($\delta=2\pi\cdot30\pm2$ Hz) which was not recorded in this experiment. The non-linear parameter, $z_{nl}$, is taken from the first experiment results and is negligible at this energy regime. Fit and data are shown in Fig. \ref{fig:nonlinear}b.

We decreased the electric field amplitude in the second experiment by a factor of 35.5 by decreasing the drive power by 31 dB. 
This expected ratio between the electric fields agrees well with the experimentally measured ratio of $E_{d,1}/E_{d,2}$=33.6$\pm$0.6. The two methods thus agree to within 5$\%$.

\section{Multi mode and energy distribution}
Thus far we treated only the case of a single mode participating in the cooling process. However, there are experimental situations in which the particle's initial energy is distributed among the three modes of the trap and moreover, this distribution is a priori unknown. 

The cooling dynamics of the three trap modes is different. First, due to the different projections of the cooling lasers along the modes axes, and second, due to the dependence of the spectrum on the modes energy distribution through micromotion \cite{Sikorsky2017}. 

\begin{figure}
	\centering
	\includegraphics[width=\linewidth, trim=3.5cm 8cm 3.5cm 8.5cm, clip]{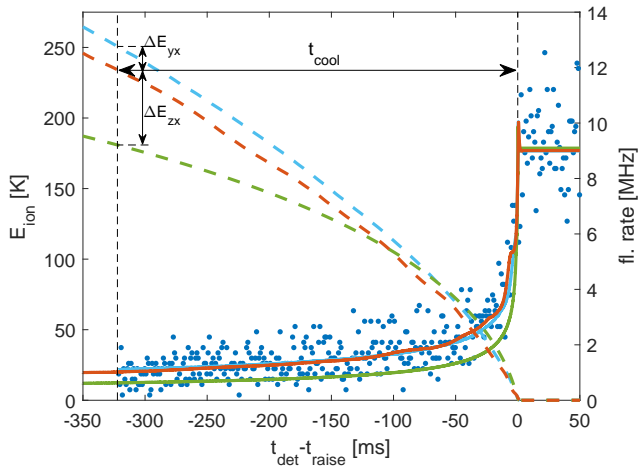}
	\caption{\textbf{Radial mode single-shot energy measurement.} Ion's secular energy (dashed lines) and fluorescence rate (solid lines) derived from the numerical solution of Eq. \ref{eq:EOMx}-\ref{eq:OBE} where the initial energy is in the x-radial mode (red), y-radial mode (light blue) and z-axial mode (green - same as in Fig. \ref{fig:singlemode}). Blue points are an example of a single experimental realization measured fluorescence of an ion prepared in the x-radial mode. The cooling time, t$_\textrm{cool}=322.2\pm1.3$ ms, translates to E$\textrm{ion}=234.3\pm0.6$ K using the x-radial mode numerically calculated curve. Errors are statistical noise which affects the fitting. If we interpret the data using the z-axial curve we underestimate the ion's energy by $\Delta\textrm{E}_\textrm{zx}\approx-53$ K. If we use the y-radial mode curve, we overestimate the ion's energy by $\Delta\textrm{E}_\textrm{yx}\approx16$ K.}\label{fig:multimode}
\end{figure}

In linear Paul traps, the axial confinement is harmonic hence the motion contains a single spectral component, usually, at a relatively low frequency of up to a few MHz. The radial motion, however, contains also fast oscillation in the 10's MHz range which is known as inherent micromotion. At large oscillation amplitudes, the spectrum significantly changes due to the appearance of side-bands \cite{Sikorsky2017,Pruttivarasin2014}. These side-bands broaden the spectrum leading to increased fluorescence. As the ion cools, the broad spectrum leads to slower cooling of the radial modes with respect to the cooling rate of the axial mode. The cooling rate as well as the fluorescence dynamics when the ion is initialized in one of the three different modes (x-red, y-light blue, and z-green) are shown in Fig. \ref{fig:multimode}. Here, the trap frequencies in the three modes were, $\omega/2\pi$=(720,980,418) kHz and the trap rf frequency was, $\Omega_\textrm{rf}/2\pi=26.51$ MHz. 

A comparison with experimental data in the radial mode is also given in Fig. \ref{fig:multimode} (blue dots). The experimental setup is the same as in Fig. \ref{fig:singlemode}, only now, we used a pulse close-to-resonance with a radial mode ($\omega_x=720$ kHz) instead of the axial mode, and a pulse time of $\sim50$ $\mu$s. We see that the experimental fluorescence agrees well with the numerically calculated curve results for an ion prepared in one of the radial modes. While the radial x- and y-modes curves are very similar, the axial mode curve is well distinguished such that a single-shot experiment is clearly sufficient to determine whether the energy was initiated in the radial or the axial modes. Below, we will show that the radial modes are also distinguishable in a single-shot, however, with reduced confidence. Interpreting the cooling signal in the wrong mode can lead to an error in the energy estimation as can be seen in Fig. \ref{fig:multimode}. 

\begin{figure}
	\centering
	\includegraphics[width=\linewidth, trim=3.5cm 8cm 3.5cm 8cm, clip]{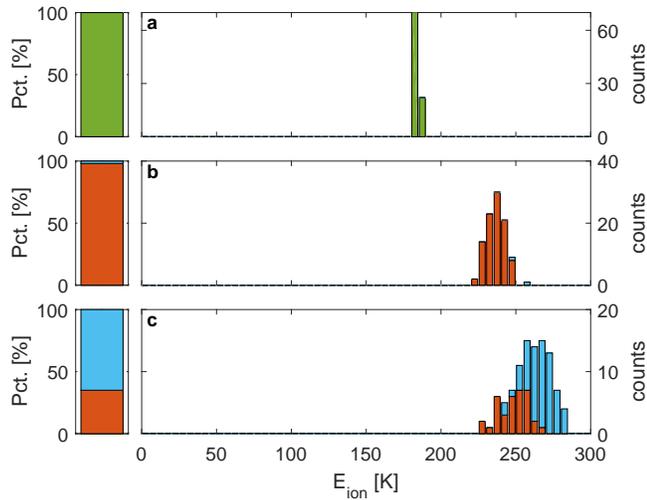}
	\caption{\textbf{Mode distinction and histogram reconstruction.} Three different experiments in which the ion is heated only in the z-axial (top - a), x-radial (middle - b) and y-radial (bottom - c) modes. We compare the experimental data to three numerically calculated curves (see Fig. \ref{fig:multimode}) and choose the model with the maximal likelihood. The histograms on the right show the ion's energy derived using the maximum likelihood model. The left bars show the initial energy mode distribution predicted by the model. Color indicates which model was used (z-green, x-red, y-light blue).}\label{fig:hisotgram}
\end{figure}

To test the confidence with which we can determine the mode in which the ion was excited we perform the following experiment: we heat the ion using close-to-resonance pulse, 100 times in each of the three different modes of the trap. We calculate, for each single-shot fluorescence data, $\mathbf{x}$, the likelihood, $\mathcal{L}$, that it corresponds to either one of the three numerically calculated curves of an ion initialized in one of the three different modes shown in Fig. \ref{fig:multimode}. Using the Poisson statistics of photon detection we can write,
\begin{equation}
    \mathcal{L}_i\left(\mathbf{x}|\theta_i\right)=\prod_t \frac{\theta_{i|t}^{x_t}e^{-\theta_{i|t}}}{x_t!}.
\end{equation}
Here, $\theta_{i|t}$ is one of the three curves ($i=x,y,z$) evaluated at the experimental cooling time, $t$, at which $x_t$ photons were measured. The mode $i$ in which the ion was initialized is determined according to a likelihood ratio test,
\begin{equation}
    \frac{\mathcal{L}_i\left(\mathbf{x}|\theta_i\right)}{\mathcal{L}_j\left(\mathbf{x}|\theta_j\right)}>\eta=1\ \forall j\neq i.
\end{equation}
Here, we choose $\eta=1$ such that type I error (experiment in mode $i$ is analyzed as $j\neq i$) and type II error (experiment in mode $j\neq i$ is analyzed as $i$) are considered on equal footing. The results are shown in Fig. \ref{fig:hisotgram}. 

The likelihood ratio test succeeded in discriminating between the axial and the radial modes for all the experimental data. None of the axial experiments were detected as radial (type I error) nor the radial experiments were detected as axial (type II error). This result is not surprising since the axial and radials models are significantly different. We bound the type I and type II error to less than $10^{-8}$ using stochastic numerical simulation analysis. 

Remarkably, when using the likelihood ratio test between the radial modes, we get 98$\%$ success when the ion is prepared in the x mode (type I error of 2$\%$) and 65$\%$ success when the ion is prepared in the y mode (type I error of 35$\%$). 
These are exceptional results considering the similarity of the x- and y-radial numerical curves as can be seen in Fig. \ref{fig:multimode} red and light solid blue lines. The expected successful discrimination between the x- and y- modes is 91$\pm$3$\%$ for 100 repetitions which is calculated using stochastic numerical simulation. Here, the error reflects the statistical noise due to the limited number of experiments. The difference between the theoretical and experimental success rate in discriminating between the modes is due to systematic error in our model or data acquisition, which biases the results.

In Fig. \ref{fig:hisotgram} we also show how a single-shot measurement is used to reconstruct the ion's energy distribution in each of the modes. We plot a histogram of the ion's energies derived from our simulated model for each of the modes. In the axial mode experiment, we see a sharp Gaussian distribution. The width of the distribution ($\sim$2 K) is attributed to slightly different cooling laser parameters in each of the experimental realization (see SM). In the radial modes, the distribution is much wider ($\sim$6.5 K for x-mode and $\sim$12.5 K for y-mode) due to the radial mode frequency variation \cite{johnson2016active} in our trap.   

\section{Conclusion}
To conclude, we show how to measure the energy of an atom in a single experimental realization and thus directly reconstruct its energy distribution. The reduced signal-to-noise due to single-shot measurement does not influence the accuracy of the method since it relies on a sharp non-linear feature of the fluorescence time dynamics.
The accuracy of the method relies on the accurate determination and control of the various experimental parameters which determine the cooling dynamics.
In return, we can use it as a precision spectroscopy tool, e.g., to measure the trap anharmonicity.
Using likelihood analysis, we can determine the distribution of energy between the modes of the trap. This opens the door for a new type of experiments, sensitive to events which re-distribute the energy between the modes without changing the total energy significantly, e.g. glancing collisions.
This method is well suited for measuring the energy release in chemical reactions involving single atoms \cite{Harter2012,Artjum_threebody,Zipkesprl,Ratschbacher2012}.

This work was supported by the Crown Photonics Center, ICore-Israeli excellence center circle of light, the Israeli Science Foundation, the U.S.-Israel Binational Science Foundation,
and the European Research Council (consolidator grant 616919-Ionology). We thank Eran Ofek and Nir Davidson for their useful remarks.


\section{Supplemental material}
\comment{
\subsection{The 64 dynamical Bloch equations}
Here we show the explicit form of Eq. 3 in the main text. We follow the derivation of \cite{Oberst1999,Sikorsky2017}. We write the density matrix operator, $\hat{\rho}$, in the interaction representation in a vector form,
\begin{equation}
	\hat{\rho}\equiv\rho_{rs}=(\rho_{11},\rho_{12},...,\rho_{87},\rho_{88}).
\end{equation}  
Here, $\rho_{ij}=\p{i}{j}$ and the energy levels are defined as (see Fig. 1 in the main text for the energy level scheme):
\begin{itemize}
	\item $\ket{1}=\ket{\level{S}{1/2}{-1/2}}$,
	\item $\ket{2}=\ket{\level{S}{1/2}{1/2}}$,
	\item $\ket{3}=\ket{\level{P}{1/2}{-1/2}}$,
	\item $\ket{4}=\ket{\level{P}{1/2}{1/2}}$,
	\item $\ket{5}=\ket{\level{D}{3/2}{-3/2}}$,
	\item $\ket{6}=\ket{\level{D}{3/2}{-1/2}}$,
	\item $\ket{7}=\ket{\level{D}{3/2}{1/2}}$,
	\item $\ket{8}=\ket{\level{D}{3/2}{3/2}}$.
\end{itemize}  
The excited state population used in Eq. 2 in the main text is defined in this notation as, $\rho_e=\rho_{33}+\rho_{44}$. Eq.3 in the main text can be written as:
\begin{equation}
	\frac{d\rho_{rs}}{dt}=\sum_{kj}L_{rs,kj}\rho_{kj}.
\end{equation}
Here, L is the 64X64 Liouville matrix given by:
\begin{widetext}
	\begin{equation}
	L=
	\begin{pmatrix}
		L_{SS|SS} & L_{SS|SP} & L_{SS|SD} & L_{SS|PS} & L_{SS|PP} & L_{SS|PD} & L_{SS|DS} & L_{SS|DP} & L_{SS|DD} \\
		L_{SP|SS} & L_{SP|SP} & L_{SP|SD} & L_{SP|PS} & L_{SP|PP} & L_{SP|PD} & L_{SP|DS} & L_{SP|DP} & L_{SP|DD} \\
		L_{SD|SS} & L_{SD|SP} & L_{SD|SD} & L_{SD|PS} & L_{SD|PP} & L_{SD|PD} & L_{SD|DS} & L_{SD|DP} & L_{SD|DD} \\
		L_{PS|SS} & L_{PS|SP} & L_{PS|SD} & L_{PS|PS} & L_{PS|PP} & L_{PS|PD} & L_{PS|DS} & L_{PS|DP} & L_{PS|DD} \\
		\hdotsfor[1]{9} \\
		L_{DD|SS} & L_{DD|SP} & L_{DD|SD} & L_{DD|PS} & L_{DD|PP} & L_{DD|PD} & L_{DD|DS} & L_{DD|DP} & L_{DD|DD} \\
	\end{pmatrix}
	\end{equation}
\end{widetext}
}
\subsection{Deriving the experimental parameters}
The numerical solution of Eq. 1-3 requires several experimental parameters which we extract from dedicated spectroscopic measurements:
\begin{itemize}
	\item We derive the trap Mathieu parameters, $a_i$ and $q_i$, from the measured trap secular frequencies, $\omega_i$, assuming $\mathbf{q}=(-q,q,0)$ and $\mathbf{a}=(-a+\tilde{a},-a-\tilde{a},2a)$. In this work, $q=0.0976$, $a=0.499\cdot10^{-3}$, $\tilde{a}=-1.254\cdot10^{-3}$. The rf frequency is $\Orf/2\pi=26.51$ MHz.
	\item We derive the projection of the lasers k-vector to the different modes using side-band spectroscopy on a quadrupole transition when the ion is ground-state cooled. The narrow linewidth spectroscopy laser at 674 nm is co-linear with the cooling and repump lasers (see Fig. 1b-c in the main text and caption for the lasers angles).
	\item We derive the magnetic field amplitude and orientation by comparing different Zeeman transitions using carrier spectroscopy on a narrow quadrupole transition. The magnitude of the magnetic field is $|B|=3\pm0.02$ Gauss and it is oriented along the imaging system axis.
	\item We measure the imaging system collection efficiency, $1/(190\pm2)$, which is defined as the number of photons collected divided by the number of photons scattered, by using the ion as a single photon source.
\end{itemize}

\begin{figure}
	\centering
	\includegraphics[width=\linewidth, trim=3.3cm 10cm 3.5cm 8cm, clip]{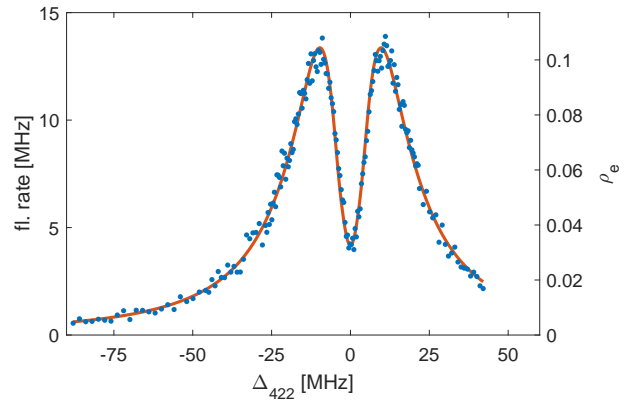}
	\caption{\textbf{Spectrum.} Blue dots are fluorescence rates measured with 25 $\mu$s detection pulses. We interlace cooling pulses between the detection pulses to eliminate mechanical (heating) effects on the spectrum. For each frequency we integrate 1,600 pulses to improve signal-to-noise ratio. Red line is the steady-state solution of Eq. 3 in the main text with the experimental parameters given in the text. 
	}\label{fig:spectrum}
\end{figure}

The ion's spectrum (Fig. \ref{fig:spectrum}), which is the steady-state solution of Eq. 3 in the main text, is mainly determined by the lasers detuning and couplings. In this experiment, we choose the lasers detuning ($\Delta_{422}/2\pi=-19$  MHz, $\Delta_{1092}/2\pi=0$ MHz) and Rabi frequency ($\Omega_{422}=0.5\Gamma_{422}$, $\Omega_{1092}=8\Gamma_{1092}$) in advance. Before the experiment is performed, we measure the ion's spectrum (blue dots in Fig. \ref{fig:spectrum}) when the ion is Doppler cooled (0.5 mK). We extract the laser parameters from a fit to the spectrum (red line in Fig. \ref{fig:spectrum}). We then servo the lasers amplitudes and frequencies using acousto-optic modulators to reach the target lasers values. We repeat this process several times until we reach convergence.

For the cooling laser, we use a noise-eating servo to keep the laser power constant during the experiment. The repump laser is well above saturation such that power fluctuations have a small effect on the spectrum. Both lasers are locked to external cavities to reduce their linewidths. For the numerical solution of Eq. 3 in the main text we use 370 kHz linewidth for both lasers. This value has a small effect on the spectrum.    

We assume linear polarization of both laser beams. The angle between the magnetic field and the polarization of the beams are, $\alpha_{422}=6^\circ$ and $\alpha_{1092}=35^\circ$.

\subsection{Energy estimation uncertainties}
To evaluate the effect of the experimental parameters on our energy estimation we look on the susceptibility of the energy to changes in these parameters.

\begin{figure}
	\centering
	\includegraphics[width=\linewidth, trim=3.8cm 8.5cm 4cm 8.5cm, clip]{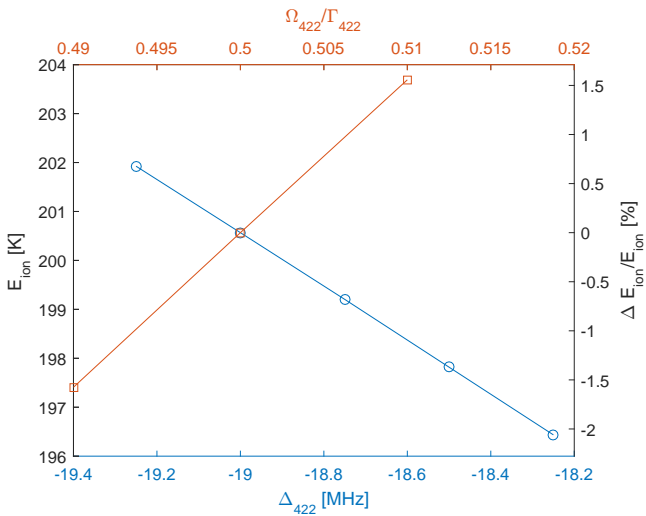}
	\caption{\textbf{Susceptibility to cooling laser detuning and Rabi frequency.} Numerical calculation of the energy for $t_{cool}$=300 ms for different cooling laser detuning, $\Delta_{422}$ (blue) and cooling laser Rabi frequency, $\Omega_{422}$ (red).}\label{fig:sus}
\end{figure}

We first check the susceptibility to the cooling beam detuning. Since we servo the spectrum only once in few hours of experiment, the cooling laser frequency can drift up to 0.5 MHz between spectrum calibrations. We change the laser cooling detuning $\Delta_{422}$ parameter and calculate the ion energy for cooling time of 300 ms. The results are shown in Fig. \ref{fig:sus} in blue color. We see that changing the laser frequency by 0.5 MHz results in relative error of 1.5$\%$ in the energy estimation. We further scan the cooling laser Rabi frequency (red). The Rabi frequency drifts about 1$\%$ between spectrum-servo scans which amounts for 1$\%$ relative error in the energy estimation. These results emphasize the need for stable laser systems for this kind of thermometry.

We now check the susceptibility to the laser beam size. In the numerical calculation, we assume that the cooling and the repump beams have a Guassian beam shape with a waist, $w_{422}$ and $w_{1092}$. For low energies below 1 K the finite beam size of the cooling and repump beams has no effect on the spectrum and the cooling dynamics. For higher energies, due to the increasing amplitude of the ion, it experiences different Rabi frequencies during its harmonic motion which becomes comparable to the beams sizes.

\begin{figure}
	\centering
	\includegraphics[width=\linewidth, trim=3.8cm 8.5cm 4cm 8.5cm, clip]{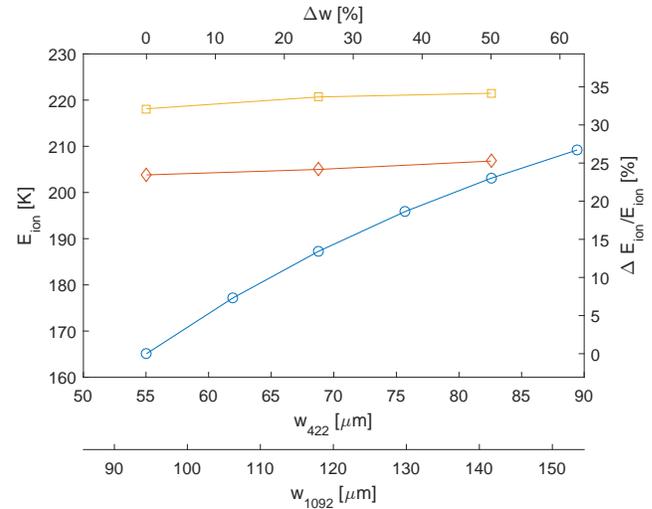}
	\caption{\textbf{Susceptibility to laser beam size.} Numerical calculation of the energy for $t_{cool}$=260 ms for different laser beam sizes, $w_{422}$ and $w_{1092}$. In Blue circles the initial energy is set to the z-mode. In yellow squares and red diamonds the initial energy is set to the y- and x-modes respectively.}\label{fig:sus2}
\end{figure}

We use the experimental results (Fig. 2, purple line in the main text) to calibrate the beam sizes, $w_{422}=55$ $\mu$m and $w_{1092}=94$ $\mu$m. These results agree well with an estimation of the cooling and repump beam sizes from the Rabi frequencies and the power of the beams measured on a power detector, $w_{422}=69$ $\mu$m and $w_{1092}=118$ $\mu$m. The 20$\%$ difference is probably due to power detector calibration and uncertainty in the vacuum chamber window reflection. In Fig. \ref{fig:sus2} we scan the beam size (blue circles) and calculate the resulting ion's energies for cooling time of 260 ms. We see that 20$\%$ error in the beam waist can lead to more than 10$\%$ error in the ion's energy. This effect is enhanced since we set the energy in the numerical calculation solely in the z-mode which is the softest mode. In yellow squares and red diamonds we set the energy in the numerical calculation in the y-mode and x-mode respectively. We see that in this case since the modes are stiffer the effect of the beam size in this energy regime is still negligible.

\comment{
\begin{figure}
	\centering
	\includegraphics[width=\linewidth, trim=3.8cm 8.5cm 4cm 8.5cm, clip]{ServoFigure.pdf}
	\caption{\textbf{Servo spectrum.} Lasers Rabi frequency ($\Omega$) normalized to the transitions linewidth ($\Gamma$) and lasers detuning ($\Delta$) during 12 hours of experimental run extracted from a spectrum fit (blue dots). After each measurement the laser power and frequency is servoed to reach the target value.}\label{fig:servo}
\end{figure}
}


\subsection{CCD thermometry modeling}
The intensity profile of an ion in a 1D classical coherent state is given by,
\begin{equation}
    I(x)=\int\limits_{0}^{2\pi/\omega} F(t)\exp(-\frac{(x-u_1\cos(\omega t))^2}{2\sigma^2})dt.
\end{equation}
Here, $u_1$ is the ion's amplitude, $\sigma$ is the width of the point-spread-function of the imaging system assumed to be Gaussian and $F(t)$ is the instantaneous fluorescence which is dependent on the velocity of the ion through Doppler shifts. We calculate $F(t)$ from the steady-state solution of Eq. 3 in the main text by modifying the laser detuning according to the instantaneous Doppler shift,
\begin{align}
    \Delta^D_{422}(t)&=\Delta_{422}-\mathbf{k}_{422}\cdot\mathbf{v}(t);\\
    \Delta^D_{1092}(t)&=\Delta_{1092}-\mathbf{k}_{1092}\cdot\mathbf{v}(t).
\end{align}
Here, $\textrm{v}(t)=-u_1\omega\sin(\omega t)$ and $\mathbf{k}=|k|\cos(46^\circ)$ for the case of an ion oscillating in the axial mode (see Fig. 1b in the main text for laser orientation).
\end{document}